# Wigner Molecular Crystals from Multi-electron Moiré Artificial Atoms

Hongyuan Li[1,2,3]†*, Ziyu Xiang[1,2,3]†, Aidan P. Reddy[4], Trithep Devakul[4], Renee Sailus[5], Rounak Banerjee[5], Takashi Taniguchi[6], Kenji Watanabe[7], Sefaattin Tongay[5], Alex Zettl[1,3,8], Liang Fu[4]*, Michael F. Crommie[1,3,8]* and Feng Wang[1,3,8]*

**Affiliations:**

[1]Department of Physics, University of California at Berkeley; Berkeley, CA, USA.

[2]Graduate Group in Applied Science and Technology, University of California at Berkeley; Berkeley, CA, USA.

[3]Materials Sciences Division, Lawrence Berkeley National Laboratory; Berkeley, CA, USA.

[4]Department of Physics, Massachusetts Institute of Technology, Cambridge, MA, USA

[5]School for Engineering of Matter, Transport and Energy, Arizona State University; Tempe, AZ, USA.

[6]Research Center for Materials Nanoarchitectonics, National Institute for Materials Science; Tsukuba, Japan.

[7]Research Center for Electronic and Optical Materials, National Institute for Materials Science; Tsukuba, Japan.

[8]Kavli Energy NanoSciences Institute at the University of California Berkeley and the Lawrence Berkeley National Laboratory; Berkeley, CA, USA.

†These authors contributed equally to this work.

*Corresponding author. Email: hongyuan_li@berkeley.edu, liangfu@mit.edu, crommie@physics.berkeley.edu and fengwang76@berkeley.edu

**Abstract:**

Semiconductor moiré superlattices provide a versatile platform to engineer new quantum solids composed of artificial atoms on moiré sites. Previous studies have mostly focused on the simplest correlated quantum solid – the Fermi-Hubbard model— where intra-atom interactions are simplified to a single onsite repulsion energy U. These studies have revealed novel quantum phases ranging from Mott insulators to quantum anomalous Hall insulators at a filling of one electron per moiré unit cell. New types of quantum solids should arise at even higher filling factors where the multi-electron configuration of moiré artificial atoms provides new degrees of freedom. Here we report the experimental observation of Wigner molecular crystals emerging from multi-electron artificial atoms in twisted bilayer $WS_2$ moiré superlattices. Moiré artificial atoms, unlike natural atoms, can host qualitatively different electron states due to the interplay between quantized energy levels and Coulomb interactions. Using scanning tunneling microscopy (STM), we demonstrate that Wigner molecules appear in multi-electron artificial atoms when Coulomb interactions dominate. Three-electron Wigner molecules, for example, are seen to exhibit a characteristic trimer pattern. The array of Wigner molecules observed in a moiré superlattice comprises a new crystalline phase of electrons: the Wigner molecular crystal. We show that these Wigner molecular crystals are highly tunable through mechanical strain, moiré period, and carrier charge type. Our study presents new opportunities for exploring quantum phenomena in moiré quantum solids composed of multi-electron artificial atoms.

**Main Text:**

Two-dimensional (2D) transition metal dichalcogenide (TMDC) moiré superlattices provide a powerful platform to simulate strongly correlated quantum solids where each moiré unit cell contains one or a few artificial atoms. Many novel quantum phases, ranging from Mott insulators(*1-3*) and generalized Wigner crystals(*4-6*) to quantum anomalous Hall insulators(*7-9*), have been observed in different TMDC moiré superlattices. Most previous studies have focused on simulating the Fermi-Hubbard model where intra-atom interactions are described by a single on-site repulsion energy $U$ that neglects the intra-atom degrees of freedom. Recent theoretical studies(*10-12*), however, have predicted that multi-electron artificial atoms in a moiré superlattice can host novel quantum states that exhibit unusual charge density distributions due to a competition between the single-particle energy level-spacing $\Delta$ and the intra-atom Coulomb repulsion energy $U$ (illustrated in Fig. 1A). Their ratio $\frac{U}{\Delta}$, usually defined as the Wigner parameter $R_W$(*13, 14*), reflects the intra-atom interaction strength. At small $R_W$ the ground state of multi-electron moiré atoms is well approximated by simply filling non-interacting orbitals in order of increasing energy. This results in a multi-electron charge distribution that peaks at the center of the moiré potential well (Fig. 1B illustrates the 3-electron case). At sufficiently large $R_W$, however, a qualitatively different electron configuration known as a Wigner molecule(*13-24*) is predicted to emerge, with electrons strongly localizing at different positions to minimize Coulomb energy (Fig. 1C illustrates the 3-electron Wigner molecule regime). This interaction-dominated electronic structure involves significant orbital reconstruction and is beyond the simplified Fermi-Hubbard model description. While spectroscopic signatures of individual Wigner molecules have previously been observed in 2D electron-gas based quantum dots(*25-31*) and short carbon nanotubes(*32, 33*), direct imaging of Wigner molecules has proved challenging. More importantly, long-range ordering of Wigner molecules into Wigner molecular crystals (a new electronic crystalline phase) has not yet been seen. Here we experimentally demonstrate the emergence of Wigner molecular crystals from multi-electron artificial atoms in twisted WS$_2$ (tWS$_2$) moiré superlattices using a new scanning tunneling microscopy (STM) imaging scheme. Many-body simulations clarify the roles played by electron-electron interactions and the moiré potential in causing Wigner molecular crystal formation.

Fig. 1D illustrates our experimental setup and device configuration. A near-60° twisted WS$_2$ (tWS$_2$) bilayer is placed on top of a 49nm thick hBN layer and a graphite back gate. We use a graphene nanoribbon array as the electrical contact to reduce contact resistance to the tWS$_2$ layer(*34, 35*). The carrier density of the tWS$_2$ is controlled by the bottom gate voltage V$_{BG}$. A bias voltage V$_{bias}$ is applied to the tWS$_2$ layer relative to the tip (which is grounded) for tunnel current measurement. Fig. 1E shows a representative STM topographic image of the tWS$_2$ moiré superlattice which exhibits a lattice constant of ~9nm (corresponding to a twist angle of ~58°). Different high symmetry stacking regions are labeled AB, B$^{S/S}$, and B$^{W/W}$, with the corresponding atomic structures illustrated in Fig. 1F. Previous studies have confirmed that the tWS$_2$ moiré superlattice creates deep moiré potential wells for electrons and holes that are localized at the B$^{W/W}$ (electrons) and AB (holes) stacking sites(*36*). This makes the tWS$_2$ moiré superlattice an ideal platform to explore multi-electron artificial atoms and Wigner molecular crystals.

One of the main challenges for studying Wigner molecules in 2D semiconductors is the tip perturbation that occurs during tunneling measurements when a strong electrical field at the tip apex can qualitatively change local electron distributions. Previously we overcame this using a sensing-layer-assisted STM technique that successfully imaged fragile TMD moiré correlated states(*4, 37*). However, the sensing layer limits spatial resolution to ~5nm which is not sufficient to resolve the internal electron configuration of moiré artificial atoms. To overcome this difficulty, we have developed a new STM scheme based on direct conduction/valence band edge tunnel current measurements.

Fig. 2A illustrates our measurement scheme for electron-doped tWS$_2$ (V$_{BG}$ > 0). Here the tip bias voltage V$_{bias}$ is tuned to the satisfy two conditions: (1) the chemical potential of the STM tip ($\mu_{tip}$) must lie within the tWS$_2$ semiconductor bandgap and (2) the vacuum energy levels of the tip and the tWS2 must roughly align. When $\mu_{tip}$ is in the semiconductor bandgap then the *doped* electrons in tWS$_2$ can tunnel into the STM tip and generate a finite current. Since this current comes from electrons near the conduction band edge, we denote it the CBE tunnel current. The CBE tunnel current directly reflects the electron spatial distribution in the tWS$_2$ layer. Because changing $\mu_{tip}$ over the entire bandgap does not alter the source of the CBE tunnel current, we can actively tune V$_{bias}$ to roughly align the vacuum energy levels of the tip and tWS$_2$. Under these conditions the electrical field between the tip and tWS$_2$ will be close to zero, thus minimizing the pertubation arising from the STM tip.

Fig. 2B shows a 2D plot of the tunnel current I-V characteristic at the B$^{W/W}$ site as a function of V$_{BG}$ and V$_{bias}$ for electron-doped tWS$_2$ (V$_{BG}$ > 0). Here the tunnel current is plotted on a log scale with different color maps for positive current (yellow) and negative current (blue). The positions of the valence and conduction band edges are labeled with red and blue dashed curves respectively (the method for identifying band edges for electron- (hole-) doped tWS$_2$ are shown in Fig. S1 (S2)). When $\mu_{tip}$ is in the semiconductor band gap (-1.3V < V$_{bias}$ < 0) a small negative tunnel current is observed that corresponds to the CBE tunnel current illustrated in Fig. 2A. The V$_{bias}$ dependence of the CBE tunnel current when $\mu_{tip}$ is in the energy gap region is characterized by a few step-like jumps. This arises from the Coulomb-induced tip perturbation, where charge accumulating at the tip apex can cause electrostatic charging or discharging of electrons in the artificial atom right below the tip. Experimentally we choose V$_{bias}$ based on the criteria of minimizing the tip perturbation in order to best facilitate the imaging of Wigner molecules in artificial atoms (see Supplementary Text section 2 and Fig. S3 for more details).

The CBE tunnel current allows us to directly image the intra-atom electron distributions within each moiré unit cell. Fig. 2C-E shows three CBE current maps for moiré artificial atoms with moiré site electron numbers of $n_e$ = 1, 2, and 3 per artificial atom. At $n_e$ = 1 (Fig. 2C, $V_{bias}$ = -0.75V, $V_{BG}$ = 2.0V) each moiré site exhibits a single peak centered at each $B^{W/W}$ site, corresponding to an electron strongly localized within the deep moiré potential. At $n_e$ = 2 (Fig. 2D, $V_{bias}$ = -0.75V, $V_{BG}$ = 4.0V) the single peak remains but is broadened due to strong intra-atomic repulsion that exceeds the single-particle gap, leading to significant orbital spreading. This orbital spreading is a consequence of interaction-induced orbital mixing and cannot be captured by a one-band Fermi-Hubbard model where interactions only alter the energetics but not the intra-atomic wavefunction.

For $n_e$ = 3 (Fig. 2E, $V_{bias}$ = -0.75V, $V_{BG}$ = 6.0V) we observe a qualitative change in the electron configuration: a trimer with three separated charge density peaks clearly emerges in each moiré artificial atom. Moreover, the electron density has a local *minimum* at the center of the artificial atom even though the moiré potential there is lowest. This is a direct visualization of a Wigner molecule, i.e., the electronic configuration where intra-atom Coulomb repulsion U dominates over the single-particle quantized energy gap $\Delta$ causing the multi-electron ground state of the artificial atom to become fundamentally different from the picture of filling "atomic" shells. This also means that single-particle orbitals at low and high energies are strongly mixed by Coulomb repulsion in multi-electron moiré artificial atoms. The moiré site electron number $n_e$ is identified through gate- and bias-dependent changes in the charge distribution current maps (see SM Fig. S3 for additional details).

We are similarly able to measure the charge distributions of hole-doped artificial atoms using a valence band edge (VBE) current measurement scheme. Fig. 2F illustrates the energy alignment diagram for STM measurement of hole-doped tWS2 where $\mu_{tWS_2}$ now lies at the VBE and the VBE tunnel current is wholly due to doped holes. The contact resistance is much larger for the hole-doped tWS2 due to a higher Schottky barrier, which leads to a non-negligible voltage drop at the electrical contact. As a result, we denote the electrical chemical potential difference between $\mu_{tip}$ and $\mu_{tWS2}$ as $V^*_{bias}$, which can be smaller than the applied $V_{bias}$ (see SM for additional details). Fig. 2G shows a 2D plot of the tunnel current I-V characteristic at the AB site as a function of $V_{BG}$ and $V_{bias}$ for hole-doped tWS2 (i.e., $V_{BG} < 0$). A nonzero VBE tunnel current is observed for $0.5V < V_{bias} < 2.5V$ when $\mu_{tip}$ is within the tWS2 semiconductor bandgap. Fig. 2H-J shows three VBE current maps for moiré artificial atoms with moiré site hole numbers of $n_h$ = 1, 2, and 3. The $n_h = 1$ map (Fig. 2H) shows a single peak centered at the AB site as expected for occupancy of a single hole. The $n_h = 2$ map (Fig. 2I), however, shows reduced charge density at the *center* of each moiré cell. Such a ring-like charge distribution is expected for a 2-hole quantum Wigner molecule. The classical 2-particle Coulomb molecule in a triangular potential well has 3-fold degeneracy (due to 3 different orientations) and the quantum Wigner molecule is expected to be in a coherent superposition of all azimuthal configurations (with appropriate weights)(*12, 13*) thus forming a ring-like pattern (see more discussion latter). The asymmetry of the rings is due to a slight uniaxial strain that breaks the C3 rotational symmetry. The $n_h = 3$ map (Fig. 2J) shows a striking array of Wigner molecules where each has a clear trimer structure consisting of three well-isolated holes. The higher clarity for hole-based Wigner molecules (Fig. 2J) compared to electron-type Wigner molecules (Fig. 2E) suggests that the Wigner parameter $R_W = \frac{U}{\Delta}$ is larger for holes than for electrons in tWS2, possibly due to a larger hole effective mass and shallower hole moiré potential.

Other parameters that significantly impact Wigner molecular crystal behavior are the moiré period and mechanical strain. Changing the moiré period alters the potential well width, thus modifying the correlation strength within a Wigner molecular crystal. Fig. 3A-C shows the evolution of the 3-electron Wigner molecule with decreasing moiré period $\lambda$ (the definition of $\lambda$ is in the SM for lattices with nonzero strain). As $\lambda$ is reduced, the trimer features become gradually weaker, and the individual electrons are no longer distinguishable by $\lambda = 8.4$nm (Fig. 3C). This is likely due to a reduction of the Wigner parameter $R_W$ as the potential well becomes narrower with decreased $\lambda$, thus leading to a reduction in band mixing and more dominant single-particle behavior.

Uniaxial strain, on the other hand, breaks the symmetry of the Wigner molecule spatial distribution. Fig. 3D-F shows 2-hole Wigner molecules as uniaxial strain is increased from $\delta = 0.01$ to $\delta = 0.26$ (see definition of $\delta$ in the SM). The ring-like structure in Fig. 3D evolves into a well-defined dimer as strain is increased to $\delta = 0.26$ (Fig. 3F). Here strain breaks the C3 symmetry of the moiré superlattice, thereby lifting the degeneracy of the 3 equivalent dimer directions and making one of them preferred as seen in Fig. 3E,F. Fig. 3G-I shows how uniaxial strain alters 3-hole Wigner molecules and leads to elongated trimer structure with a modified internal particle distribution as strain is increased from $\delta = 0.03$ to $0.39$. Fig. 3J-L shows how uniaxial strain alters 4-hole Wigner molecules as strain is increased from $\delta = 0.03$ to $0.39$. The unusual pattern of 4-hole Wigner molecule at low strain matches well with the recent theoretical prediction reported in Ref (*12*). Similar strain-modified Wigner molecule behavior is also observed for electron-type Wigner molecules (see SM Fig. S4 for more details).

The Wigner molecular crystals we observe experimentally can be understood through a theoretical model that takes into account electron-electron interaction strength, moiré periodicity, and strain. Here we focus on the hole-doped regime where the essential physics is captured by an effective continuum Hamiltonian for layer-hybridized $\Gamma$ valley holes(*38, 39*):

$$H = \sum_i \left\{\frac{p_i^2}{2m} + \Delta(r_i)\right\} + \sum_{i<j} V_c(r_i - r_j).$$

Here $\Delta(r) = -2V_0 \sum_{n=1}^{3} \cos(\boldsymbol{g}_n \cdot \boldsymbol{r} + \phi)$ is the first-harmonic moiré potential with $\boldsymbol{g}_n = \frac{4\pi}{\sqrt{3}\lambda}\left(\sin\frac{2\pi n}{3}, \cos\frac{2\pi n}{3}\right)$, $V_c(r) = \frac{e^2}{4\pi\epsilon r}$ is the inter-particle Coulomb interaction potential, and $m$ is the effective mass. The moiré potential has minima in the AB stacking regions and can be expanded thereabout as $\Delta(r) \approx \frac{1}{2}kr^2 + c_3 r^3 \cos(3\theta) + \cdots$ where $k = \frac{16\pi^2 V_0 \cos(\phi)}{\lambda^2}$ and $c_3 = \frac{16\pi^3 \sin(\phi)}{3^{3/2}\lambda^3}$, thus giving rise to effective "moiré atoms" with three-fold anisotropy. The physics of the Wigner molecular crystal is controlled by three length scales(*10*): (i) the quantum confinement length $\xi_0 = \left(\frac{\hbar^2}{mk}\right)^{\frac{1}{4}}$ which determines the spread of the one-electron wavefunction, (ii) the Coulomb confinement length $\xi_c = \left(\frac{e^2}{16\pi\epsilon k}\right)^{\frac{1}{3}}$ which determines the equilibrium separation of two point-charges in a harmonic well, and (iii) the moiré period $\lambda$. The Wigner molecular crystal regime is characterized by a hierarchical relationship between these length scales: $\xi_0 < \xi_c < \lambda$.

We model the ground state charge density configuration of Wigner molecular crystals using two complementary approaches: self-consistent Hartree-Fock (HF) theory for the full continuum model, and exact diagonalization (ED) for single moiré atoms. HF accounts for the full moiré superlattice but treats electron interactions in mean-field, while ED treats electron correlations exactly but is restricted to a single artificial atom. The continuum model parameters we have chosen correspond to a moiré atom energy level spacing $\Delta \equiv \hbar\omega \approx 37 meV$ (here $\omega = \sqrt{\frac{k}{m}}$ is the oscillator frequency) and intra-moiré-atom interaction energy $U \equiv \frac{e^2}{4\pi\epsilon\xi_0} \approx 190$ meV roughly consistent with previous works(*37, 40*) and yielding $R_W \approx 5.1$ (see SM for further details). Fig. 4A shows the resulting theoretical hole density obtained by HF for $n_h = 3$. The calculation yields a Wigner molecular crystal in good agreement with the experimental data of Fig. 2J. Fig. 4B shows the HF hole density at $n_h = 2$ in the presence of a uniaxial strain of $\delta = 0.1$ which is seen to stabilize a dimer configuration that is also consistent with the experimental data of Fig. 3F. However, we note that the HF results tend to spuriously break rotation symmetry of the two-hole molecules: it yields a dimer even without strain (see more discussion in Methods of the SM). This drawback can be overcome with the ED calculation.

      The ED charge density for a single unstrained moiré atom with 3 holes is seen in Fig. 4C and shows good overall agreement between the two numerical models and the experimental data. The unstrained $n_h = 2$ ED charge density of Fig. 4D, on the other hand, shows a doughnut-like configuration that is more symmetrical than the data of Fig. 2I, 3D. After adding a small amount of strain to the ED calculation, however, the charge density relaxes in the strain direction and becomes more prolate (Fig. 4E), consistent with the experimental data of Fig. 2I, 3D. The $n_h = 2$ configuration is thus seen to be highly sensitive to strain while also exhibiting an inherently quantum delocalization effect arising from quantum fluctuations of the three-fold degenerate classical dimer configurations. A well-defined dimer configuration emerges in the ED charge density (Fig. 4F) with additional uniaxial strain similar to the experimental data of Fig. 3E,F. These calculations highlight the dominant role of intra-moiré electron-electron interactions that lead to Wigner molecule physics at fillings $n > 1$ in a TMDC moiré superlattice. This is in strong contrast to the Fermi-Hubbard model where such interactions only alter the energetics and not the intra-atomic wavefunction. It is also useful to notice that the ED charge density (Fig. 4C-F) is more diffuse than the HF results (Fig. 4A,B) since the HF calculation includes Coulomb repulsion between neighboring Wigner molecules that is not included in the EF calculation and

which partially counteract the intra-Wigner-molecule repulsion and shrinks the Wigner molecule size(*41*).

In conclusion, we observe the emergence of Wigner molecular crystals in moiré artificial atoms using a new STM tunnel current measurement scheme. This novel state of matter represents a new type of electron crystalline phase that arises from multi-electron artificial atoms and has no analog in conventional quantum solids made from natural atoms. We show that Wigner molecular crystals can be manipulated by changing charge carrier type, moiré period, and mechanical strain in a moiré heterostructure. The intra-atom electron correlations within Wigner molecular crystals provide new opportunities to explore spin, charge, and topological phenomena that are distinct from conventional quantum solids.

**Acknowledgments:**

**Funding:** This work was primarily funded by the U.S. Department of Energy, Office of Science, Basic Energy Sciences, Materials Sciences and Engineering Division under Contract No. DE-AC02-05-CH11231 within the van der Waals heterostructure program KCFW16 (device fabrication, STM spectroscopy). Support was also provided by the National Science Foundation Award DMR-2221750 (surface preparation). S. T. acknowledges primary support from U.S. Department of Energy-SC0020653 (materials synthesis), NSF CMMI1825594 (NMR and TEM studies on crystals), NSF DMR-1955889 (magnetic measurements on crystals), NSF ECCS2052527 (for bulk electrical tests), DMR 2111812, and CMMI 2129412 (for optical tests on bulk crystals). K.W. and T.T. acknowledge support from the JSPS KAKENHI (Grant Numbers 21H05233 and 23H02052) and World Premier International Research Center Initiative (WPI), MEXT, Japan. The work at Massachusetts Institute of Technology was supported by the Air Force Office of Scientific Research (AFOSR) under award FA9550-22-1-0432.

**Author contributions:** H.L., M.F.C., and F.W. conceived the project. H.L. and Z.X. fabricated the heterostructure device. H.L. and Z.X. performed the STM/STS measurement, H.L., Z.X., A.Z., M.F.C. and F.W. discussed the experimental design and analyzed the experimental data. A.P.R., T.D., and L.F. performed the theoretical calculations. R.S., R.B. and S.T. grew the $WS_2$ crystals. K.W. and T.T. grew the hBN single crystal. All authors discussed the results and wrote the manuscript.

**Competing interests:** Authors declare that they have no competing interests.

**Data and materials availability:** The data supporting the findings of this study are included in the main text and in the Supplementary Information files and are also available from the corresponding authors upon request.

**Supplementary Materials**

Materials and Methods

Supplementary Text

Figs. S1 to S9

Videos S1 and S2

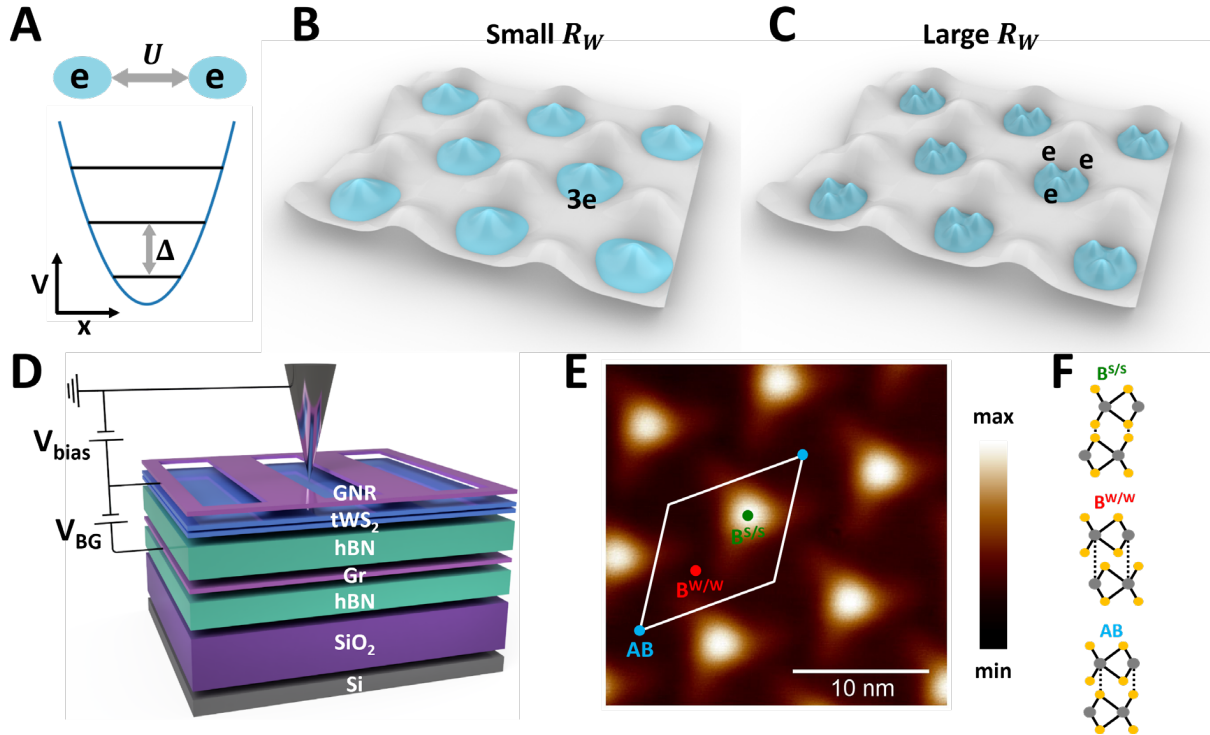

**Fig. 1. Multi-electron artificial atoms in a moiré superlattice.** (**A**) Schematic of the intra-atom Coulomb repulsion energy $U$ (top) and energy gap $\Delta$ between single-particle states in a parabolic potential well (bottom). The Wigner parameter $R_W$, defined as $\frac{U}{\Delta}$, reflects the internal interaction strength. (**B,C**) Illustration of an array of moiré artificial atoms with three electrons per moiré unit cell under different $R_W$ values. (**B**) For small $R_W$ the three electrons in each artificial atom center around the potential minimum due to the six-fold-degenerate single-particle ground states. (**C**) For large $R_W$ a triangular Wigner molecule charge configuration emerges in each artificial atom, generating a Wigner molecular crystal in the overall moiré superlattice. (**D**) Schematic of STM measurement setup for a gate-tunable near-58-degree twisted WS$_2$ (tWS$_2$) moiré superlattice. tWS$_2$ is placed on top of a 49nm thick hBN layer and a graphite substrate that defines the back gate. A back gate voltage V$_{BG}$ is applied to control the charge carrier density in the tWS$_2$. A tip bias V$_{bias}$ is applied between the tWS$_2$ and STM tip to induce a tunnel current. A graphene nanoribbon (GNR) array is placed on top of tWS$_2$ as the contact electrode. (**E**) A typical STM topographic image of the tWS$_2$ surface allows the identification of different high-symmetry stacking regions (AB, B$^{W/W}$, B$^{S/S}$). V$_{bias}$ = -3V, I = 11pA, V$_{BG}$ = -5V. (**F**) Atomic structure of the three different stacking regions in a moiré unit cell.

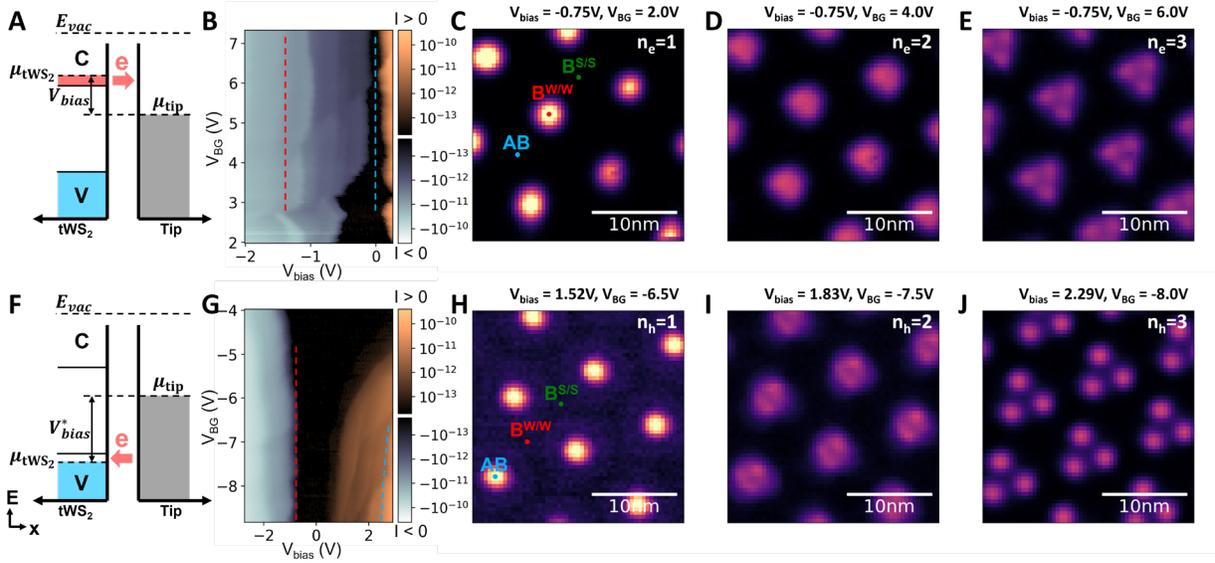

**Fig. 2. CBE/VBE tunnel current measurement of Wigner molecules.** (**A**) Schematic energy diagram for conduction band edge (CBE) tunnel current measurement of electron-doped tWS$_2$. The tWS$_2$ chemical potential $\mu_{tWS_2}$ is near the conduction band edge. When the tip chemical potential $\mu_{tip}$ (controlled by V$_{bias}$) is aligned within the band gap of the tWS$_2$, the tunnel current arises from the doped electrons at the CBE. V$_{bias}$ is tuned to roughly align the vacuum energy levels of the tip and tWS$_2$ so that the electric field near the tip apex is minimized. (**B**) Tunnel current I-V characteristic as a function of V$_{BG}$ measured at the B$^{W/W}$ site for electron-doped tWS$_2$. The current is plotted on a log scale and the positive (yellow) and negative (blue) parts use different color maps. The valence and conduction band edges are labeled with red and blue dashed curves. (**C-E**). CBE tunnel current maps of artificial atoms with different electron numbers (n$_e$): (**C**) n$_e$ = 1 (V$_{bias}$ = -0.75V, V$_{BG}$ = 2.0V), (**D**) n$_e$ = 2 (V$_{bias}$ = -0.75V, V$_{BG}$ = 4.0V), (**E**) n$_e$ = 3 (V$_{bias}$ = -0.75V, V$_{BG}$ = 6.0V). (**F**). Schematic energy diagram for the valence band edge (VBE) tunnel current measurement of hole-doped tWS$_2$. $\mu_{tWS_2}$ is near the valence band edge for hole-doped tWS$_2$. When $\mu_{tip}$ is aligned within the band gap of tWS$_2$ the tunnel current arises from doped holes at the VBE. The gap between $\mu_{tip}$ and $\mu_{tWS2}$ is denoted as $V^*_{bias}$ instead of $V_{bias}$ since the real bias on the tunneling junction $V^*_{bias}$ is lower than $V_{bias}$ due to the large contact resistance. (**G**). Tunnel current I-V characteristic as a function of V$_{BG}$ measured at the AB site for hole-doped tWS$_2$ plotted similar to (**B**). Note that the valence band edge (labeled with red dashed curve) is not located at V$_{bias}$ = 0 due to the large tip perturbation at negative V$_{bias}$ (see SM for additional details). (**H-J**). VBE tunnel current maps of artificial atoms having different hole numbers (n$_h$): (**H**) n$_h$ = 1 (V$_{bias}$ = 1.52V, V$_{BG}$ = -6.5V), (**I**) n$_h$ = 2 (V$_{bias}$ = 1.83V, V$_{BG}$ = -7.5V), (**J**) n$_h$ = 3 (V$_{bias}$ = 2.29V, V$_{BG}$ = -8.0V).

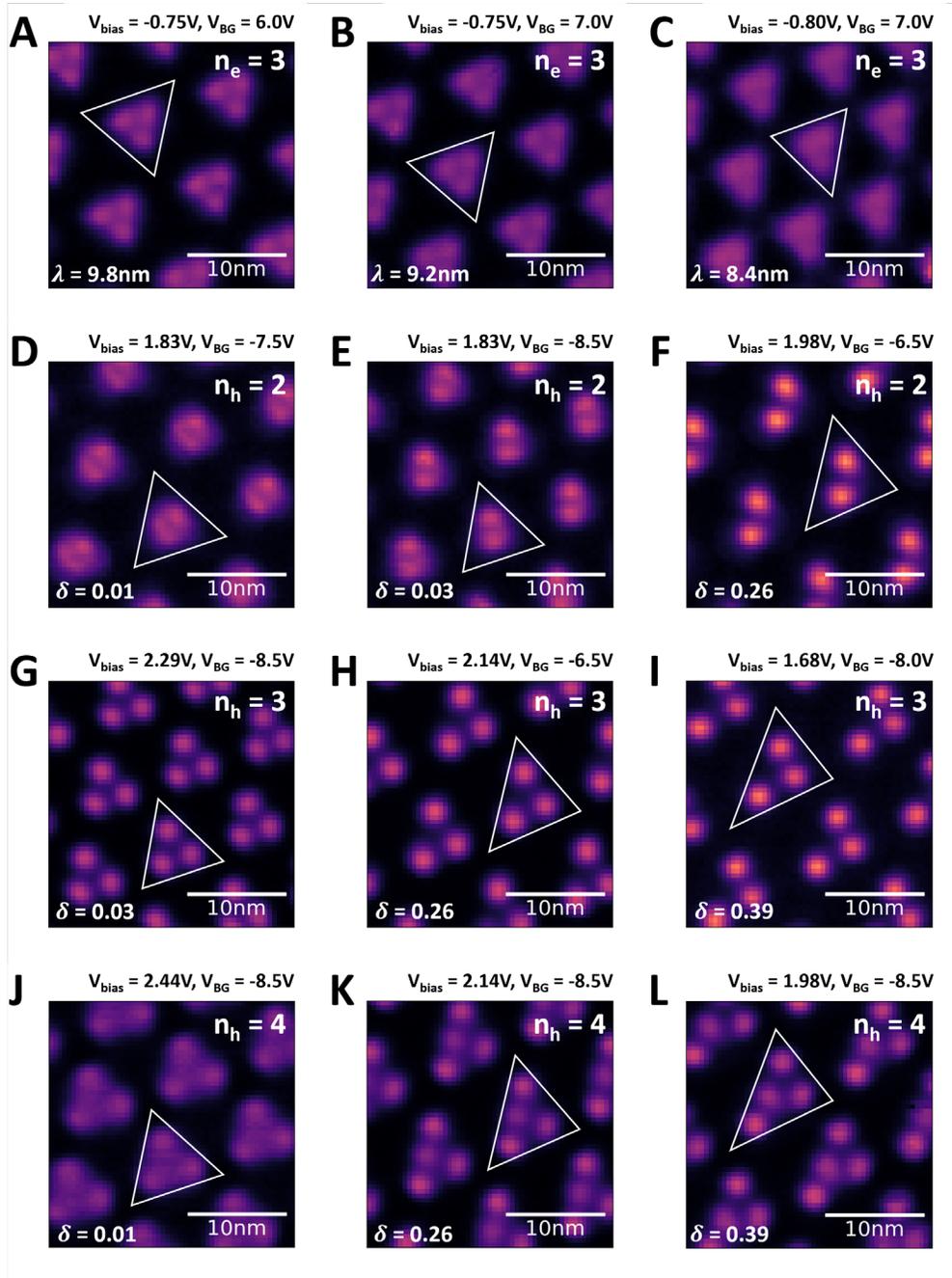

**Fig. 3. Configuration engineering of Wigner molecular crystal.** (**A-C**) Evolution of 3-electron Wigner molecules as the moiré period is decreased from (**A**) $\lambda = 9.8$nm to (**B**) $\lambda = 9.2$nm, and to (**C**) $\lambda = 8.4$nm. (**D-F**) Evolution of 2-hole Wigner molecules as uniaxial strain is increased from (**D**) $\delta = 0.01$ to (**E**) $\delta = 0.03$, and to (**F**) $\delta = 0.26$. (**G-H**). Evolution of 3-hole Wigner molecules as uniaxial strain is increased from (**G**) $\delta = 0.03$ to (**H**) $\delta = 0.26$, and to (**I**) $\delta = 0.39$. (**J-L**). Evolution of 4-hole Wigner molecules as uniaxial strain is increased from (**J**) $\delta = 0.01$ to (**K**) $\delta = 0.26$, and to (**L**) $\delta = 0.39$. (**A**) and (**D**) are reproduced from Fig. 2E and Fig. 2I respectively. Exact definitions of $\lambda$ and $\delta$ can be found in the SM. The white triangle in each pannel labels the potential well contour.

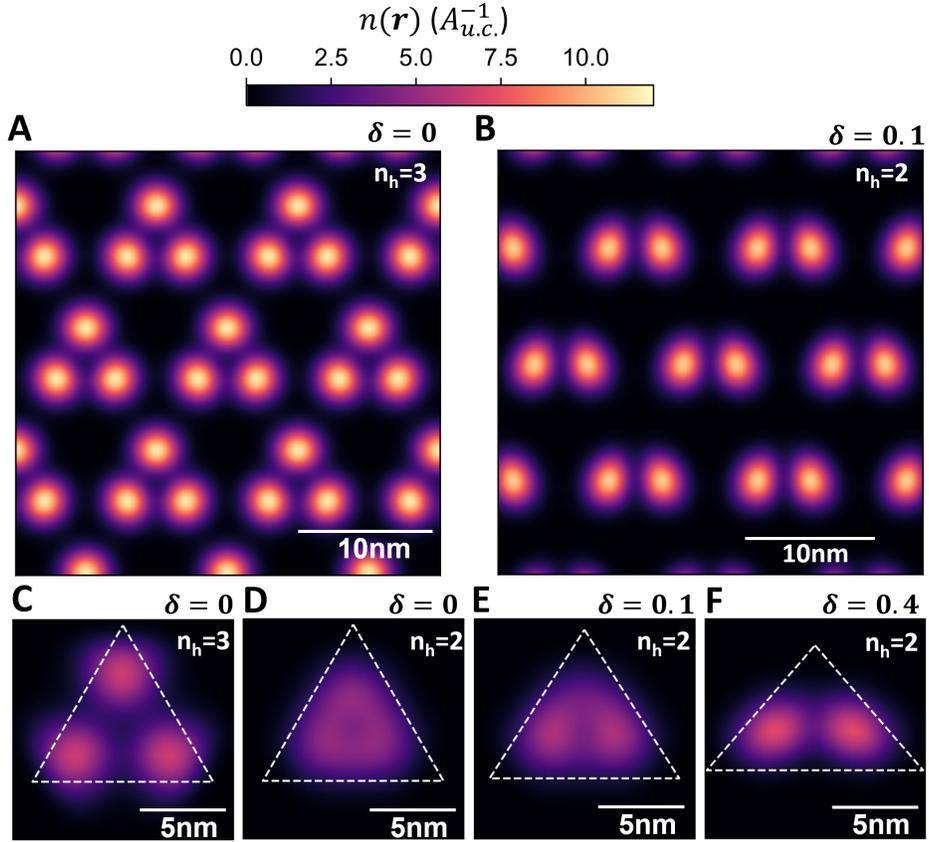

**Fig. 4. Numerical simulations of Wigner molecular crystal.** (**A**) Self-consistent Hartree Fock (HF) ground state hole density map for a continuum model with $n_h = 3$ (assuming full spin/valley polarization). (**B**) HF ground state hole density map for $n_h = 2$ with $S_z=0$ and applied strain ($\delta = 0.1$, see the SM for definition). (**C**) Exact diagonalization (ED) ground state for 3 fully spin polarized electrons in a single moiré potential well. (**D-F**) ED ground state of 2 electrons in a singlet state with different values of strain: (**D**) $\delta = 0$, (**E**) $\delta = 0.1$, and (**F**) $\delta = 0.4$. The white dashed triangles in (**C-F**) label the potential well contour. $A_{u.c.}$ is the moiré unit cell area. The HF (**A,B**) and ED (**C-F**) results share the same color bar.

18